\documentclass{kluwer}    
\usepackage{epsfig}
\newdisplay{guess}{Conjecture}
\def\sma{$\cal M_\odot$}

\begin{document}                                                                
\begin{article}
\begin{opening}         
\title{Starburst dwarfs - fueling and morphological evolution} 
\author{Nils \surname{Bergvall$^1$, G\"oran \"Ostlin$^2$, Josefa Masegosa$^3$, 
Erik Zackrisson$^1$}}  
\runningauthor{Nils Bergvall}
\runningtitle{Starburst dwarfs}
\institute{1) Astronomical observatory, Box 515, S-75120 Uppsala, Sweden;
 2) Institut d'Astrophysique de Paris, 98bis Boulevarrd Arago, F-75014 Paris, 
France; 3) Instituto de Astrofisica de Andalucia, CSIC, Apdo. 3004, 18080, 
Granada, Spain}
\date{Sept 10, 1999}

\begin{abstract}
The effects of mergers in low mass galaxies are poorly understood. In this
paper we analyze different observational evidences which support the view
that mergers can trigger starbursts in dwarf galaxies. 
We discuss the relationships between blue compact galaxies (BCGs), dIs, dEs and 
low surface brightness galaxies (LSBGs) and present some evidences 
which favour that strong starbursts are caused by mergers between dEs and LSBGs.
\end{abstract}

\end{opening}           

\section{Introduction}

Since Zwickys \cite{zwicky} early discovery of Blue Compact Galaxies (BCGs) a 
large amount of work has been devoted to investigate their nature.  
Major questions are what causes the burst, how long it
can be sustained and if the heating and mechanical energy it produces eventually 
will cause the cas to be expelled from the system. Recent models (Mac Low and 
Ferrara \cite{mac}) indicate that the galaxy probably can retain its gas if 
$\cal M \ge $10$^7$ \sma. Cooling and reaccretion may later
ignite a  new burst and in a Hubble time a few such bursts may occur. Another 
possibility is the young  galaxy hypothesis.  Today  we  have two  hot
candidates, IZw18 and SBS0335-052, both with  extremely low
metallicities. The possible presence of old stars is still debated however 
(Aloisi et al. \cite{aloisi}; Hunter \& Thronson \cite{hunter}; Papaderos et al. 
\cite{papaderos1}). In both cases  signs of merger
activity are evident.  This also seems  to  be a  common  property of  the 
most  active starburst dwarfs and may be the most 
likely scenario to explain the burst activity.

\section{Morphology and colours. The starburst progenitors}

Different classification 
systems have been proposed for starburst dwarfs. Telles et al. \cite{telles} 
separate 
two different types, 
one regular type and the other more luminous, having an irregular structure. 
These may be related, modified in their morphology by evolution.

In the merger scenario one of the progenitors has to be a gas rich, metal poor 
galaxy, i.e. a dI galaxy, a Low Surface Brightness Galaxy (LSBG) or an isolated 
HI cloud. We have studied the global properties 
of a few LSBGs and BCGs to look for possible relationships and find that the 
global parameters like disk scale length
and central surface brightness are quite similar. It has been claimed (Papaderos 
et al. \cite{papaderos2}) 
that LSBGs cannot be related to BCGs since these two parameters differ 
significantly. However, the problem may be related to the lack 
of very deep images. The scalelengths
are derived from regions too close to the central starburst. Fig.  1 from a  
deep  B image  of ESO  338-IG04 shows that the slope of the profile changes at 
$\mu _B\sim$27$^m$ arcsec$^{-2}$.  An
exponential fit to  the outer  profile results in a central surface brightness 
of $\mu _{B,0}\approx$24.5,  typical of  a faint  LSBG. The
difference  between a  power  law fit  and  an exponential  fit is
however minor and we need to look  at the colours of the halo to
determine what type  of  galaxy  may  be
involved. Fig.  2 shows results from surface  photometry in the
optical/near-IR of the BCG halos  and the global colours of our sample
of  blue LSBGs  compared  to  predictions from  our  spectral  evolutionary
models. For comparison the colours of the halo of IZw18 (\"Ostlin et al., in 
prep.) is also indicated. All LSBGs are starforming so the short burst model 
does not apply. Thus the ages are $\ge$ 2-3 Gyr. Although we cannot go as far 
out into the halo in the near-IR as in the optical, we find that the halo 
colours of the BCGs range from those of LSBGs to those of very old red stellar 
populations. One  of the galaxies, ESO 338-IG04, has an exponential LSB disk and 
at the same time  contains a system of
globular  clusters  (\"Ostlin et al. \cite{ostlin1}) with  a  full  
range  of ages  and
indications of a  few bursts in the past. The GCs indicate the presence of a 
relatively massive, very old system. A probable scenario is therefore a merger 
between an LSBG and a dE/dSph. Other strong supports  of merger  activities come 
from  a study  of the H$\alpha$
velocity  field in this galaxy and a  handful of other luminous BCGs (\"Ostlin 
et al. \cite{ostlin2}).

\begin{figure}
\centerline{\epsfig{file=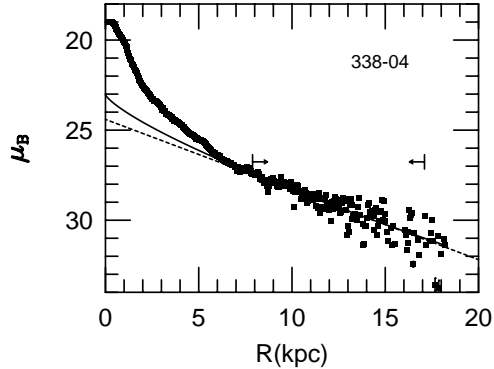, width=16pc}}
\caption{The luminosity profile of ESO 338-IG04 in B, corrected for an 
inclination of $\it i$= 62$^o$. The hatched line is a pure exponential fit and 
the full drawn line is the best fit to a Sersic law with an exponent of 0.8. 
(ESO NTT.)}
\end{figure}

\begin{figure}
\centerline{\epsfig{file=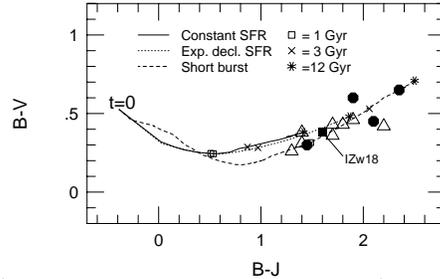, width=16pc}}
\caption{Optical/near-IR colours of our blue LSBGs (triangles) and the halos of 
the 4 luminous BCGs ESO 338-IG04, ESO 350-IG38, ESO 400-G34 and ESO 480-IG12 
(filled dots) and IZw18 (filled square). The predicted colour evolution of three 
different star forming scenarios is displayed: constant star formation rate 
(SFR), an exponentially declining SFR and a short initial burst. Data obtained 
mainly at ESO.}
\end{figure}

\section{Metallicities and HI content}

If mergers are the cause of luminous starbursts, why would they be fed by LSBGs 
and not dIs? Bergvall et  al. \cite{bergvall}
argued  that LSBGs are more attractive due to  the constraints set by  the 
metallicities and the HI contents of dwarf galaxies in general. Under the 
assumption that the samples available in the literature are representative, the
luminosity-metallicity  diagram indicates  a  luminosity  difference
between a  BCG and a  dI {\sl at the same metallicity} of  $\sim$ 2
magnitudes. A bursting dI would obtain a $\cal M_{HI}$/L$_B$ drastically
lower than what is observed for BCGs. LSBGs have no such problem. Fig. 3
illustrates the  situation in a slightly different way. There
seems to be a clear distinction  between dI/Is and LSBGs in the O/H -
$\cal M_{HI}$ diagram. It shows very nicely that they are the same family, the 
LSBGs being the less evolved and more gas rich ones. If we ignite a starburst, 
the position of a galaxy would shift towards a lower gas mass and a higher 
metallicity. This indicates that the best guess progenitors for the burst 
component of the
BCGs are the LSBGs. In the most luminous cases, e.g. ESO 338-04 and ESO 400-43, 
as for SBS0335, LSBGs seems to be the only option.

The two most metal poor starburst dwarfs, IZw 18 and SBS0335-052, are located in 
the 
same region as the LSBGs. Are these two galaxies in an early phase of the 
merging process? The disturbed HI morphology
of IZw 18 (van Zee et al. \cite{vanzee}) may point to this. The morphology of 
the HII/HI cloud of SBS0335-052 (see also Pustilnik et al 1999, in preparation) 
also opens the long standing question of the low metallicity. Are we 
measuring the metallicity of the nearly pristine HI gas cloud illuminated by
the starburst or has it something to do with the underlying galaxy?

A possible scenario that would explain the different properties of dIs
and BCGs  is that the star  formation efficiency is ruled  by the dark
matter potential field.  The DM is more centrally  concentrated in dEs
than in dIs and this may make the difference.  Since the DM dominates
the mass distribution in dwarf  galaxies, the burst we see in luminous
BCGs  today therefore mimics  the first  burst in  the history  of the
dE.  A merger between an LSBG  and a  dI would probably be less dramatic.

\begin{figure}
\centerline{\epsfig{file=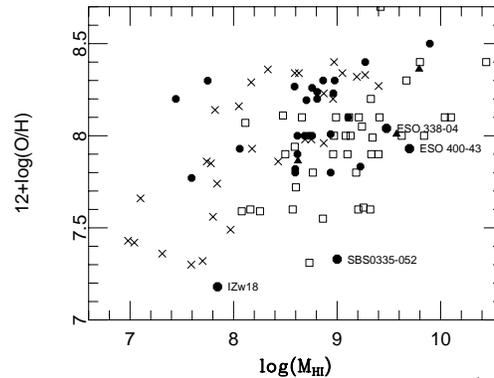, width=16pc}}
\caption{HI mass versus oxygen abundances for a sample of dIs (stars), BCGs 
(filled dots) and LSBGs (squares) obtained from the literature, including our 
own data from ESO, Nancay and Parkes.}
\end{figure}

\end{article}
\end{document}